\documentclass[pra]{revtex4}
\usepackage{graphicx}

\begin{document}
\title{Visualizing Teleportation}
\author{Scott M. Cohen}
\email{cohensm@duq.edu}

\affiliation{Department of Physics, Duquesne University,
Pittsburgh, Pennsylvania 15282}
\affiliation{Department of Physics, Carnegie-Mellon University,
Pittsburgh, Pennsylvania 15213}

\begin{abstract}
\noindent A novel way of picturing the processing of quantum information is described, allowing a direct visualization of teleportation of quantum states and providing a simple and intuitive understanding of this fascinating phenomenon. The discussion is aimed at providing physicists a method of explaining teleportation to non-scientists. The  basic ideas of quantum physics are first explained in lay terms, after which these ideas are used with a graphical description, out of which teleportation arises naturally.
\end{abstract}

\maketitle

\section{Introduction}
One of the most exciting and fastest-growing fields of physics today is quantum information. Especially since the discovery by Shor \cite{ShorFactor,Gerjuoy} that there exist calculations for which a quantum computer is apparently far more efficient than a classical computer, interest in understanding quantum information has increased at an impressive rate. One widely publicized discovery that has emerged from work in this field is teleportation \cite{BennettTele}. While not precisely equivalent to the process enjoying widespread fame amongst fans of Star Trek (``Beam me up, Scotty"), the phenomenon referred to here is nonetheless fascinating, and perhaps even astonishing. The reason for the widespread publicity of this rigorously proven (and experimentally tested \cite{ExptTele1,ExptTele2,ExptTele3,ExptTele4,ExptTele5}, though not yet unambiguously demonstrated) scientific prediction is almost certainly in large part due to the fact that it shares the same name as the just-mentioned, intriguing idea from science-fiction.

The usual way of describing teleportation is through mathematical equations, and this mathematics is relatively straightforward, as has been amply demonstrated elsewhere \cite{NielsenChuang,BennettTele}. Hence, an understanding of this phenomenon is accessible to physicists, other scientists, and those possessing a reasonably strong level of mathematical skill. There does, on the other hand, seem to be a good deal of misunderstanding of teleportation amongst non-scientists, with the notion floating around that the amazing phenomenon shown regularly in episodes of Star Trek -- that is, of material objects being teleported from one place to another -- has actually turned out to be possible in real life. Nothing could be further from the truth, of course, so we are left wondering how to rectify this unfortunate state of affairs. The question I address here is the following: can the true (scientific) phenomenon of teleportation be understood by others, those without much skill in mathematics? The usual explanations will certainly fail in this regard, even if carefully presented by a competent physicist, because mathematics has a well-known tendency to scare people away, and in any case, the mathematics of teleportation is not all \textit{that} simple. The paper is addressed to physicists possessing a solid understanding of quantum physics (including graduate students), with the aim to provide a method by which such a physicist can explain teleportation to someone who is not mathematically inclined. Thus, the objective is ultimately, though indirectly, to educate the general public about teleportation, and by extension, quantum mechanics itself. The approach involves only the most basic ideas about quantum physics, and while it does not entirely avoid mathematical expressions, it uses only the simplest mathematics (one only needs to accept that certain objects are either $0$ or $1$) and relies almost entirely on ``pictures", allowing the layperson to visualize -- and thus, understand -- what is happening.

In the following sections, I will describe my method of directly visualizing teleportation. These sections are written as if addressed to the layperson. The next section explains the probabilistic nature of quantum physics by considering ``quantum coins", which are examples of two-level systems. This section describes how one should think about measurements, what is meant by probabilities for classical systems, and then how these ideas can be used to describe quantum systems. Then, in Section~\ref{vqip}, I present my graphical approach to understanding the dynamics of quantum information processing, which is then used in Section~\ref{tele} to explain in pictures how teleportation of quantum states is possible. One of the crucial observations will be that a shared entangled state on, say, systems $a$ and $b$, provides the parties with multiple ``images" of the state of an additional system $A$. The ability to manipulate these images -- independently by each party, and differently from one image to the next -- is what allows teleportation to be accomplished. More generally, these ideas provide important insights into why entanglement is a valuable resource, as I have described in detail elsewhere, and they have been useful in understanding other aspects of quantum information processing \cite{inpressNote,ourNLU}.

\section{Probabilities}
Perhaps the most fundamental aspect of quantum theory is that it can only make predictions in terms of probabilities. In general even if one has a complete description of the state of a quantum system, one will not know ahead of time what the outcome of a given measurement will be. This is in direct contradiction with our everyday experience, which we refer to as ``classical". For example, a flipped classical coin which lands heads (``heads" is then a complete description of the state of this coin), is known with certainty to be heads, and also with certainty to \textit{not} be tails. That is, if we know the state of a classical coin (in this case ``heads"), we can predict with certainty the answer to any reasonable question we choose to ask (or ``measure") about that coin (for example, ``Is it tails?"). We therefore need to understand what is meant by the ``state" of a quantum system and how this state relates to probabilities and outcomes of measurements. The following definition of a measurement will be adequate for our purposes.

\textbf{Definition:} A \textit{measurement} is a procedure that provides answers to a collection of yes-no questions, which is both mutually exclusive (when the answer to one of the questions is ``yes", the answer to all the others is ``no") and complete (all possibilities are included; that is, one of the questions will always be answered in the affirmative). The single question that receives the ``yes" answer is referred to as the \textit{outcome} of the measurement.

For example, since a classical coin is either heads or tails, and these two possibilities are mutually exclusive, a measurement on a classical coin is a procedure that answers the two questions ``Is it heads?" and ``Is it tails?" Since the coin will always be one or the other, there will always be a ``yes" answer to one of these questions, and then the other question is always answered ``no". Hence these two questions do indeed constitute a measurement according to the above definition. If ``Is it heads?" is answered affirmatively, then ``heads" is the outcome of the measurement.

It turns out that these two questions also constitute a measurement on quantum coins. However, in contrast to the classical case in which this is the \textit{only} possible measurement, there is a vast array of possible measurements on quantum coins. This will become clearer from the discussion in the following sections, where we introduce a compact way of describing these things, a way commonly used in quantum mechanics.

\subsection{Classical coins and classical probabilities}
Consider again a flipped classical coin. The coin lands either heads or tails. It will be useful to use a somewhat abbreviated notation: $|H\rangle$ for heads and $|T\rangle$ for tails. The statement that ``if it is heads, it is not tails" (that is, has zero probability of being tails) will be represented as
\begin{eqnarray}\nonumber\nonumber
	\langle T|H\rangle = 0.
\end{eqnarray}
The left-facing bracket $|H\rangle$ represents the known initial state (``It is heads.") and the right-facing bracket $\langle T|$ represents the question (``Is it tails?"). The number ($0$) appearing on the right-hand side of the equal sign then gives the probability that with this initial state, the answer to this question will be yes. For the above example, we have that the probability is $0$, which is as expected since when the coin is $H$ it will never be $T$. Note that it is useful to use the left- and right-facing brackets, so that we can easily read off what is the initial state and what is the question being asked about it. Simply writing $TH=0$ in the above equation would lead to confusion when we discuss two coins (see below), which might have an initial state where one is tails, the other heads, represented by $|TH\rangle$.

Perhaps an even more trivial statement ``if it is heads, then it is heads" (with certainty, or with probability one), will similarly be represented as
\begin{eqnarray}\nonumber
	\langle H|H\rangle = 1.
\end{eqnarray}
Again, the right-facing bracket contains the question $\langle H\vert$, or ``Is it heads?", and the fact that the expression is equal to $1$ indicates that the answer to this question will \textit{always} be ``yes" when the initial state is $\vert H\rangle$. These statements are trivial because if we know the state of a classical coin, we can predict with certainty whether it will be heads or tails when we look at it.

Although the remaining equations will look a bit more involved, the only mathematics the reader need understand is contained in the above two equations, along with two others that are almost exactly the same. The discussion in the remainder of this paper will follow from the four simple statements, $\langle H|H\rangle = 1$, $\langle T|T\rangle = 1$, $\langle T|H\rangle = 0$, $\langle H|T\rangle = 0$.

Next let us consider two coins. In this case, a complete list of mutually exclusive possibilities is $HH,~HT,~TH,~TT$. We can make statements in exactly the same way we did above, for example ``if they are $HH$, then they are not $HT$", which in our notation is written
\begin{eqnarray}\nonumber
	\langle H_1T_2|H_1H_2\rangle = \langle H_1|H_1\rangle\times\langle T_2|H_2\rangle = (1)\times(0) = 0,
\end{eqnarray}
where the subscripts ($1,2$) have been inserted for clarity to indicate which coin is which. Note that in this equation, we have equated the expression $\langle H_1T_2|H_1H_2\rangle$ with the product of two expressions, $\langle H_1|H_1\rangle$ and $\langle T_2|H_2\rangle$. This is because any question about the two coins jointly is the same as two questions, one about each of the coins separately.

It is obviously also true that ``if they are $HH$, then they are $HH$", so
\begin{eqnarray}\nonumber
	\langle H_1H_2|H_1H_2\rangle = \langle H_1|H_1\rangle\times\langle H_2|H_2\rangle = (1)\times(1) = 1.
\end{eqnarray}
For three coins, there are eight possibilities ($HHH,~HHT,~HTH,~THH,~HTT,~THT,~TTH,~TTT$) and the same notation will readily account for this case, as well. We will not need to consider more than three coins here, though it is in principle straightforward to do so.

\subsection{Quantum coins and quantum probabilities}
Quantum coins behave very differently as compared to their classical counterparts, and quantum probabilities must be understood in very different ways. We still have heads and tails, $|H\rangle$ and $|T\rangle$, as possible states of a quantum coin. We refer to these two states as being ``orthogonal" to each other, by which we simply mean that they are mutually exclusive: if the quantum coin is $H$, it is definitely (with certainty) not $T$, and vice-versa. We note that the four equations appearing in the previous section are equally true for both quantum and classical coins. However, there now exist some very strange possibilities. If I were to suggest that a classical coin can be both $H$ and $T$ at one and the same time, you would be completely justified in thinking I'd gone slightly crazy. I am going to tell you, though, that at least in a certain (though very real) sense, this is the case for quantum coins (though you may still wonder a bit about my sanity). The point is that, in the quantum case, it makes complete sense to ask questions such as: ``If the coin is $H$, is it half $H$ and half $T$?"; or we can turn this around and ask ``If the coin is half $H$ and half $T$, is it $H$?" Neither of these questions makes any sense whatsoever when referred to a classical coin. On the other hand, for a quantum coin these are not only legitimate questions, but they are in fact very important ones (we do not consider the negligible possibility of a classical coin landing on its edge, and in any case this bears no relationship to what we mean by a quantum coin being half $H$ and half $T$).

To represent these questions, we can write the state ($Q$) of a quantum coin that is half $H$ and half $T$ as
\begin{eqnarray}\nonumber
	|Q\rangle = \frac{1}{2}|H\rangle + \frac{1}{2}|T\rangle.
\end{eqnarray}
Then the answer to the question, ``If the coin is half $H$ and half $T$, is it $H$?" is answered by the equation,
\begin{eqnarray}
	\langle H|Q\rangle & = & \langle H|~\left(~\frac{1}{2}|H\rangle + \frac{1}{2}|T\rangle~\right)\nonumber \\
				& = & \frac{1}{2}\langle H|H\rangle + \frac{1}{2}\langle H|T\rangle\nonumber \\
				& = & \frac{1}{2}(1) + \frac{1}{2}(0) = \frac{1}{2}\nonumber,
\end{eqnarray}
which should be interpreted as meaning ``yes, with probability $1/2$", implying also ``no, with probability $1-1/2=1/2$" [In quantum mechanics, it is actually the square of the object on the left-hand side of the foregoing equation that represents the probability, rather than that object itself, which is known as the ``probability amplitude"; however, although the difference between probabilities and probability amplitudes is crucial to the understanding of quantum mechanics, I have chosen in the present discussion to overlook this distinction for the benefit of the layperson to whom these ideas are aimed, as they would only serve to complicate matters, causing unnecessary confusion amongst the intended audience]. The left-facing bracket $|Q\rangle$ represents the known initial state, and the right-facing bracket $\langle H|$ represents the question (``Is it heads?"). The number $1/2$ appearing on the right-hand side of the last line then gives the probability that with this initial state, the answer to this question will be yes. The point to understand here is that even though we have a complete description ($Q$) of the state of the quantum coin, we do not generally know in advance whether the coin will be $H$ or $T$ when we look at it. We can only predict in terms of probabilities: if we perform this experiment many times, half the time the answer will be yes and the other half of the time it will be no. Furthermore, there are many more questions we can ask in the quantum, as compared to the classical, case. We are no longer restricted to asking ``is the coin $H$?" or ``is it $T$?", but we can ask other questions, such as the reverse of the question we just answered,
\begin{eqnarray}\nonumber
	\langle Q|H\rangle & = & \left(\frac{1}{2}\langle H| + \frac{1}{2}\langle T|\right)|H\rangle\nonumber \\
				& = & \frac{1}{2}\langle H|H\rangle + \frac{1}{2}\langle T|H\rangle\nonumber \\
				& = & \frac{1}{2}(1) + \frac{1}{2}(0) = \frac{1}{2}\nonumber.
\end{eqnarray}
We see that the question ``If the coin is $H$, is it half $H$ and half $T$?" has the same answer as the previous question: ``yes, with probability $1/2$; and no, with probability $1/2.$"

We note that in the remainder of the paper, instead of phrasing questions as ``is the coin half $H$ and half $T$?", we instead ask whether it is ``equal parts" $H$ and $T$. While there is no real difference between these two questions, this rephrasing allows us to simplify the notation by dispensing with the factors of $1/2$ that have appeared in the above discussion. In doing so, the equations will not yield the same numbers as probabilities for the various questions, but this will not hamper the presentation since the numerical values of the probabilities are not crucial to the ideas we wish to convey: we just need to remember that certain objects are equal to $1$ and others are equal to $0$.

\section{Teleportation}\label{vqip}
What exactly do we mean by teleportation in the context of quantum information? It is not a material object that is being teleported, but rather the state of a quantum system. We will assume that the system is a quantum coin, with a complete set of mutually exclusive (orthogonal) states being ``heads" and ``tails", which we may denote as $|H\rangle$ and $|T\rangle$. Suppose Alice and Bob are physicists in locations widely separated from each other. They each have a quantum coin -- labeled $a$ and $b$, respectively -- and these two coins are in the state
\begin{eqnarray}\nonumber
	|B_0\rangle_{ab} = |H_aH_b\rangle + |T_aT_b\rangle,
\end{eqnarray}
where the subscripts used here refer to system \textit{a} (\textit{b}) in Alice's (Bob's) possession. This state of two quantum coins has a very strange property, which is known as entanglement, and the state itself is an example of a maximally entangled state. Entanglement is a rather strange sort of correlation between quantum systems, which manifests itself in the state $B_0$ by the fact that neither system \textit{a} nor \textit{b} can be considered to have a definite ``state of its own" independent of the other system: whatever is the state of coin $a$, coin $b$ will have the same state, but one cannot say anything about the state of either coin independent of the other one. It is this property of entanglement that is credited with enabling Alice and Bob to accomplish teleportation.

Alice is given another coin (system \textit A), prepared in a state
\begin{eqnarray}\nonumber
	|S_A\rangle=c_H|H_A\rangle+c_T|T_A\rangle
\end{eqnarray}
with arbitrary coefficients $c_H$ and $c_T$ that are completely unknown to her and to Bob. If $c_H=1/2$ and $c_T=1/2$, we have the case discussed in the previous section, where the coin is equally likely to be found to be $H$ or $T$. For other values of these coefficients, the two possibilities will in general not be equally likely. Alice's task is to perform operations on the systems in her possession (\textit a and \textit A) in such a way that Bob will end up with his system (\textit b) in precisely the state $|S_b\rangle$, which is the same state as $|S_A\rangle$, but now on the distant system \textit b. It turns out that this task can be accomplished if Alice communicates information to Bob (perhaps via a telephone) about what she ended up doing to her systems, after which Bob performs a rather simple quantum operation, dependent on the information obtained from Alice, on system \textit b.

An important point to understand in what follows is that nothing either of them does in this process provides even the slightest information about the coefficients $c_H$ and $c_T$, so the state ($S$) that has been teleported remains completely unknown to the parties. This aspect of teleportation becomes even more amazing if one considers the amount of information that is conveyed: the information contained in a quantum state is far greater than the amount actually transmitted from Alice to Bob via the telephone (as we will see below, the amount transmitted via the telephone is two classical bits, enough to convey which one of four possibilities has been chosen). True, the $\textit{classical}$ information one can encode in a two-level quantum system cannot exceed one bit (one bit is the amount of information needed to choose between two possibilities, such as $|H\rangle$ and $|T\rangle$). But if Alice wanted to tell Bob how to create the state in his own lab by communicating with him over a phone line, this would require an $\textit{infinite}$ amount of classical information; that is, enough information to completely describe the arbitrary numbers, $c_H$ and $c_T$ (it is infinite because one of these numbers might well be an irrational number such as $\pi$, having a decimal expansion that is unending, never repeating itself). Of course, Alice and Bob are both \textit{completely ignorant} of what these numbers are, so even if it were possible to transmit an infinite amount of information, they don't even know what information they would need to send! Nonetheless, when they share entanglement, it is possible for the two of them, by working together, to create the unknown state on Bob's coin $b$ with the communication of only two classical bits.

\subsection{Visualizing quantum information processing}
Let us now introduce the pictorial method which will be used to visualize teleportation. The simple diagrams we will use to depict states of multiple quantum coins, held by two different parties, are familiar to many researchers working in quantum information. We will now illustrate how these diagrams are used to represent quantum states, and then how they can be used to follow what happens to these coins when measurements are performed by one of the parties. Then, we will be ready to use them for visualizing teleportation.
\subsubsection{States of quantum coins}
To depict the state of a single quantum coin labeled $A$ (standing for Alice; she will also have the other coin labeled $a$, while Bob's single coin is labeled $b$), we may use a simple box diagram,
\begin{eqnarray}\nonumber\label{fig:Eq3}
\centering
\includegraphics{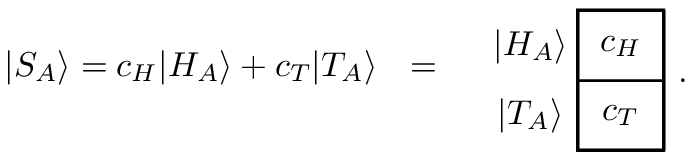}
\end{eqnarray}
The coefficients $c_H$ and $c_T$ appearing in the boxes indicate ``how much" is in that part of the state $S_A$ of coin $A$. The next example illustrates the case where there are two coins ($A$ and $b$) held by two different parties. Then, the state of these two coins might be
\begin{eqnarray}\nonumber\label{fig:Eq4}
\centering
\includegraphics{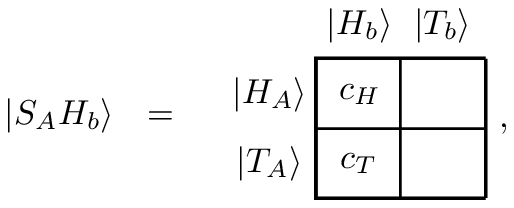}
\end{eqnarray}
with $S_A$ as given above. The empty squares on the right-hand side of this diagram represent the fact that system \textit b is ``not $T$" (has zero probability of being tails); the $c_H$ in the upper-left corner represents the probability the coins are both heads; and the $c_T$ in the lower-left, the probability Bob's coin is heads and Alice's is tails.

If there are three parties involved, a three-dimensional cube could be used to represent this situation. However, it will serve our present purposes to represent both of Alice's systems along the vertical dimension of the diagram. We might have coins $A$ and $b$ as in the previous example, and coin $a$ being heads, the overall state of these three coins represented as
\begin{eqnarray}\nonumber\label{fig:Eq5}
\centering
\includegraphics{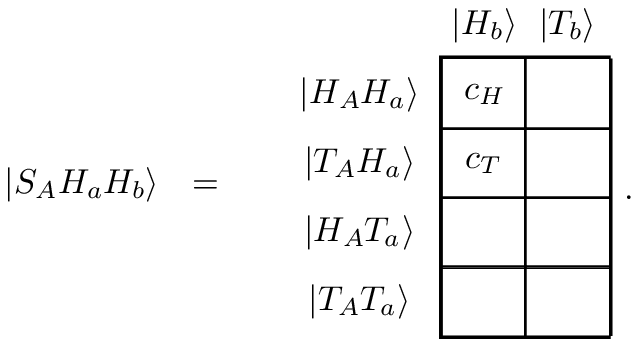}
\end{eqnarray}
If instead the \textit{a,b} systems are both $T$, this picture is
\begin{eqnarray}\nonumber\label{fig:Eq6}
\centering
\includegraphics{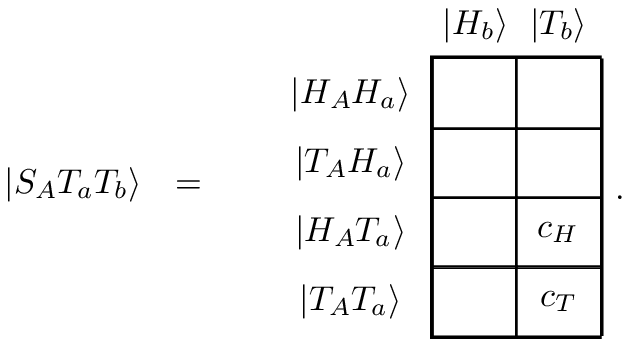}
\end{eqnarray}
Now consider what happens if we add the previous two equations together. Then our two coins \textit{a,b} are ``equal parts in $HH$ and in $TT$", which is what we previously referred to as the ``maximally entangled state" $|B_0\rangle_{ab} = |H_aH_b\rangle + |T_aT_b\rangle$. The corresponding diagram looks like
\begin{eqnarray}\nonumber\label{B0_dgrms}\label{B0SA}
\centering
\includegraphics{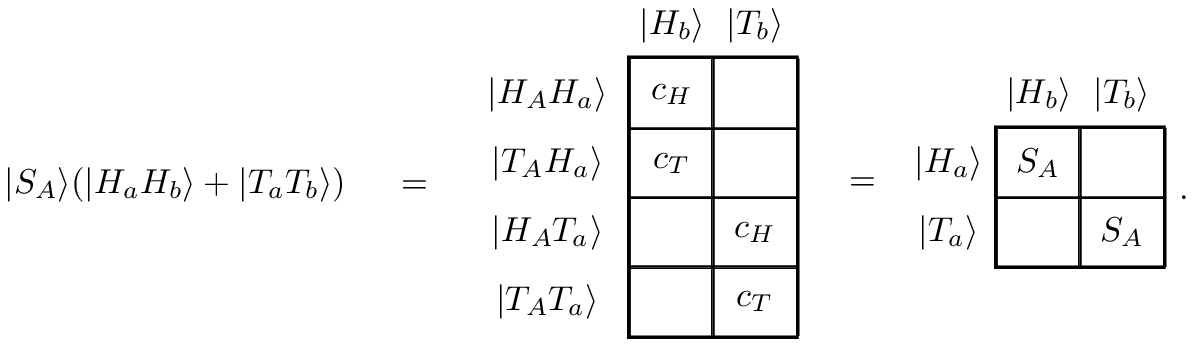}
\end{eqnarray}
Notice how there are now two images of the state $|S_A\rangle$. This observation turns out to be rather useful in understanding entanglement \cite{inpressNote,ourNLU}, but we will not need to discuss such issues here. Let us now look at how to represent measurements by use of these diagrams.

\subsubsection{Measurements on quantum coins}
Suppose Alice and Bob share three quantum coins in the state represented in the last equation of the previous section
, and Alice wants to know something about her coins. If she measures coin $a$ and discovers it is $H$, then we have
\begin{eqnarray}\nonumber\label{fig:Eq8}
\centering
\includegraphics{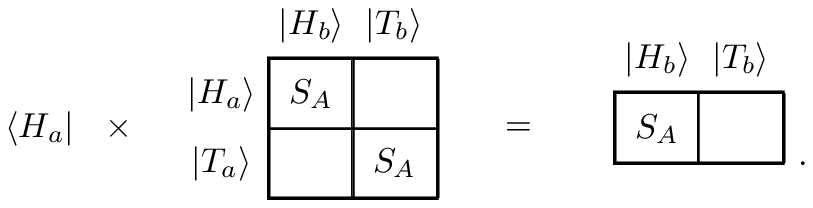}
\end{eqnarray}
Recall that when the right-facing bracket $\langle H_a|$ is attached to the left-facing one $|H_a\rangle$ on the left of this equation, we get $\langle H_a|H_a\rangle = 1$, which ``preserves" the upper row, whereas $\langle H_a|T_a\rangle=0$, indicating that the bottom row is annihilated (multiplied by $0$), which is why it no longer appears on the far right of this equation. The interpretation is as follows: when the question ``Is coin $a$ heads?" is answered in the affirmative the other coins are left in the state $|S_AH_b\rangle$. We see how this measurement acts on \textit{both} of the images simultaneously, rather than on the two independently. The upper-left image has been preserved intact, but the other image was annihilated, disappearing altogether.

On the other hand, if the outcome of Alice's measurement had been that coin $a$ was $T$, this would be represented as
\begin{eqnarray}\nonumber\label{fig:Eq9}
\centering
\includegraphics{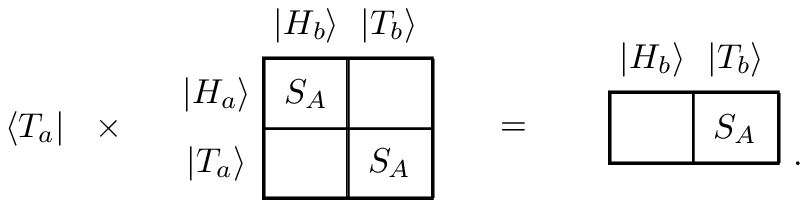}
\end{eqnarray}
In this case, the upper-left image has disappeared and the one in the lower-right has been preserved intact. In each of these cases, the state of coin $A$ is unchanged, but that of coin $b$ is left in a state that corresponds directly to the outcome of Alice's measurement on $a$. If she discovers that coin $a$ was $H$ (or $T$), then coin $b$ ends up $H$ (or $T$).

Alternatively, she could do a measurement that includes the question ``Is coin $a$ equal parts $H$ and $T$?" If the answer to this question is yes, then
\begin{eqnarray}\nonumber\label{fig:Eq10}
\centering
\includegraphics{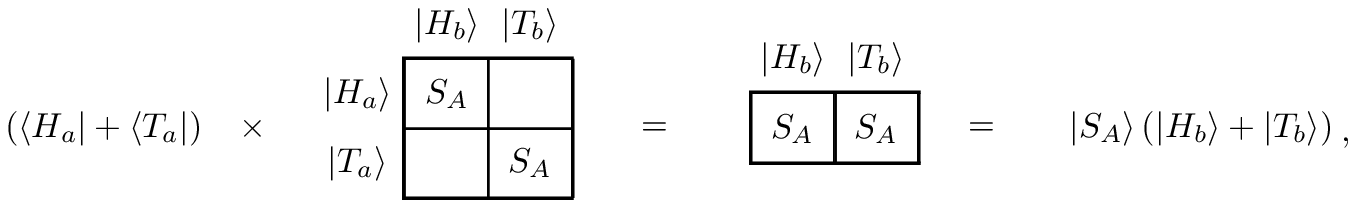}
\end{eqnarray}
which is just a sum of the previous two equations (notice how after each of the three measurement outcomes we have just considered, the two images have been collapsed into a single row). Once again we see that the state of coin $b$ ends up corresponding to the outcome of Alice's measurement on coin $a$. This illustrates some of the strangeness that resides in entangled states of quantum systems: no matter what measurement Alice makes on coin $a$ and no matter what outcome she obtains from that measurement, the resulting state of coin $b$ will correspond directly to that outcome.

The way the images of $S_A$ appear in the diagram is crucial. The fact that the two start out in different rows \textit{and} in different columns will be important in what is to come. If entanglement between systems \textit{a,b} was absent, for example if they were in the (unentangled) state $\left(|H_a\rangle +|T_a\rangle \right)|H_b\rangle$, then this would be represented by (recall that $|S_A\rangle = c_H|H_A\rangle + c_T|T_A\rangle$)
\begin{eqnarray}\nonumber\label{fig:Eq11}
\centering
\includegraphics{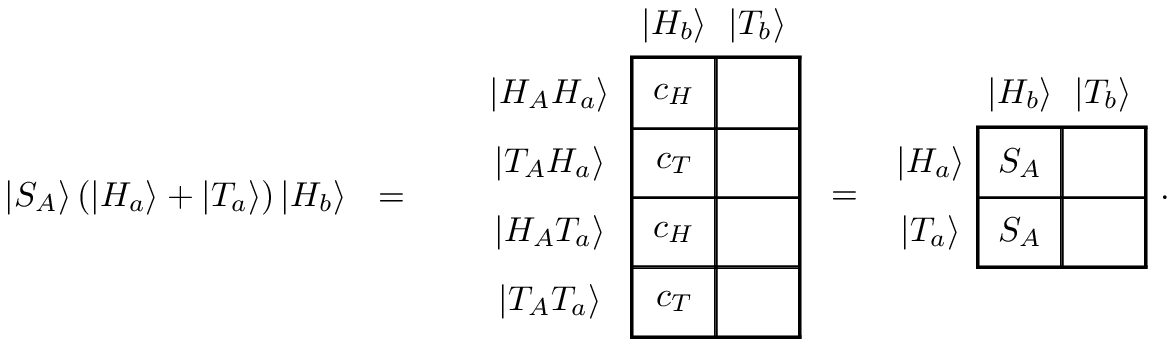}
\end{eqnarray}
Under these circumstances, Bob's view of the lower image of $S_A$ is ``obstructed" by the presence of the upper image; the two images effectively appear as one to him. As will become clear in the following section, the presence of entanglement between the \textit{a,b} coins will be necessary for them to accomplish teleportation. We will see that it is Bob's (and Alice's) ability to ``see" the two images separately, and the consequent ability for each of them to act \textit{differently} on one of the images as compared to the other, that is crucial to their success.

In the next section, we turn to the task of teleporting the state $S_A$ onto Bob's coin $b$. To begin this process, Alice will perform a measurement that asks ``joint" questions; that is, questions about both coins in her possession simultaneously. As an example, she could ask if they are both $H$. That is,

\begin{eqnarray}\nonumber\label{fig:Eq12}
\centering
\includegraphics{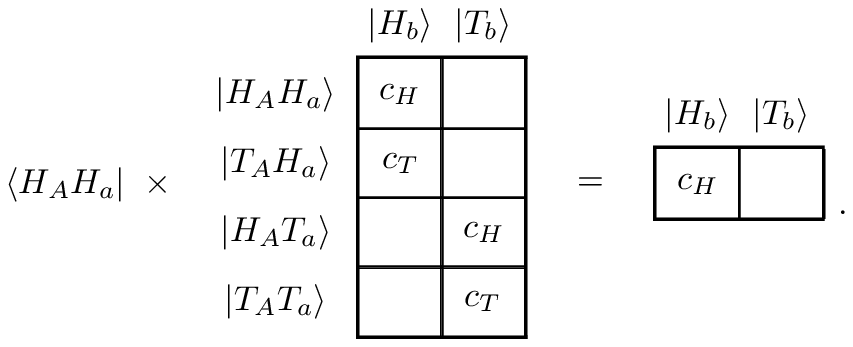}
\end{eqnarray}
The $c_H$ appearing in the box on the right corresponds to the probability that the answer to this question will be ``yes". More important for our purposes is to recognize that when this is the outcome of the measurement, coin $b$ ends up $H$, once again a consequence of the initial entanglement between coins $a$ and $b$. 

Now let us see how teleportation is possible.

\subsection{Visualizing teleportation}\label{tele}
Teleportation is accomplished with the aid of the extra systems $a,b$ in the entangled state $|B_0\rangle_{ab}$. System $A$ starts in state $|S_A\rangle$, discussed above, and this is the state they will teleport. Alice will ask a set of joint questions, which together constitute a measurement, about the state of the two coins in her possession, $a$ and $A$. The first question she asks is whether these two coins are equal parts $HH$ and $TT$. When the answer is yes, we have

\begin{eqnarray}\nonumber\label{fig:Eq13}
\centering
\includegraphics{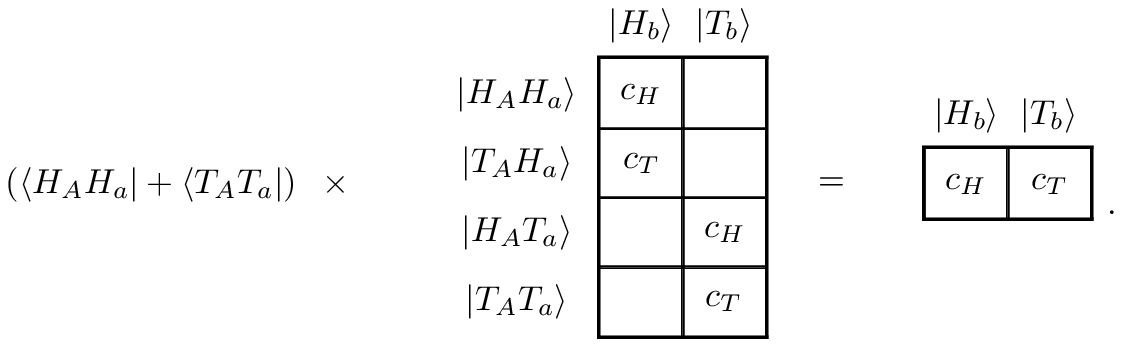}
\end{eqnarray}
Notice how the middle two rows are annihilated by this outcome (because these rows correspond to a situation where the two coins are different -- one $H$ and one $T$ -- whereas we are asking if they are the same), and the remaining rows are collapsed into a single row. Now, if we look carefully (or perhaps, not even so carefully) at the final diagram in this picture, we will arrive at a rather startling conclusion. We see that the state of Bob's system \textit b is now $|S_b\rangle = c_H |H_b\rangle + c_T|T_b\rangle$. That is, the unknown state $|S_A\rangle$, originally on system \textit A, is now on Bob's system \textit b. Furthermore, the question asked by Alice had nothing whatsoever to do with the coefficients $c_H$ and $c_T$, which determine what the original state of coin $A$ was. Hence, the parties remain completely ignorant of the state $S$, yet that state has been successfully teleported!

We are not quite finished, however, since we would like for Alice and Bob to be able to teleport no matter which joint question ends up being the outcome of Alice's measurement. Because of the probabilistic nature of the quantum world, she cannot choose the outcome of her measurement. Instead, Alice effectively asks all of the questions in her chosen measurement and then must wait for Nature to decide which question she (Nature, that is) will choose as the outcome. The nice thing about Nature is that she will tell Alice which question was chosen.

There must be four questions in a complete set of questions making up a joint measurement on coins $A,a$. Let me illustrate with one other question how Alice and Bob can succeed with teleportation, and then the reader is asked to believe that they can also succeed with either of the remaining two questions (these can be treated in a very similar way to the one shown here \cite{minusNote}). The second question is: Are coins $A,a$ equal parts $TH$ and $HT$? The corresponding diagram is

\begin{eqnarray}\nonumber\label{fig:Eq14}
\centering
\includegraphics{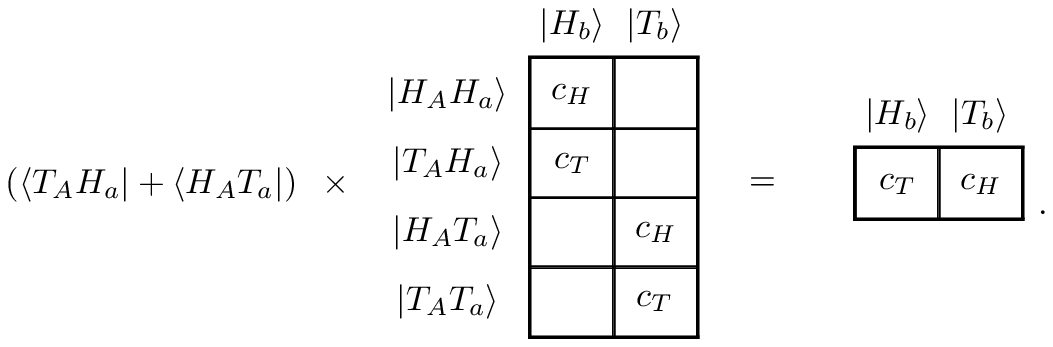}
\end{eqnarray}
Here, the first and last rows are annihilated by this outcome, and the middle two are collapsed into a single row. Looking at the final diagram, we see that coin $b$ is left in the state $c_T|H_b\rangle+c_H|T_b\rangle$, which has the coefficients $c_H$ and $c_T$ exchanged in comparison to the state $S$ that we want it to be in. However, Alice knows that this is the question to which Nature answered yes, and she can call Bob on the telephone and inform him of this fact. Once he knows this is the question that was answered affirmatively, all he needs to do is ``flip his quantum coin". Recall that this is a quantum coin, which he cannot simply pick up and turn over. Instead what we mean by this is that he exchanges $H$ for $T$ and vice-versa. In turns out that this is a legitimate action that can be performed on a quantum coin, and it will leave coin $b$ in the desired state: $c_T|H_b\rangle+c_H|T_b\rangle \rightarrow c_T|T_b\rangle+c_H|H_b\rangle=|S_b\rangle$. Notice also that they again remain completely ignorant of the coefficients $c_H$ and $c_T$ -- nothing that has happened has provided them with any such information, nor have they needed it. It turns out that no matter which of the four outcomes of Alice's measurement she obtains, once she informs Bob of that outcome, he will be able to perform a legitimate action on his quantum coin that will leave his coin in the state $|S_b\rangle$. All Bob needs to know, in order to choose which action to perform, is the outcome of Alice's measurement: Alice needs only to send him two bits of information, enough to choose between one of the four possible outcomes. Furthermore, none of the four outcomes provides either party with any information about the coefficients, $c_H$ and $c_T$, so they both remain completely ignorant of the original state they have just successfully teleported.

The diagrams provide a great deal of insight into what is going on. The crucial observation is the presence of two images of the state $S$, resulting from the entanglement between coins $a,b$. Alice does a measurement that, while not acting independently on these two images, does act \textit{differently} on them, as we alluded to above. This measurement picks out \textit{different} parts of $S$ from the different images in a way such that all of $S$ is preserved and none of it is repeated. For example, in the previous example, the $H$ part is preserved from the lower-right image, and the $T$ part from the upper-left one (and vice-versa in the example before that). This suggests (and it is indeed the case) that for coins that have more than two sides (a six-sided quantum die, for example), the parties can teleport the state of such an $n$-sided coin by sharing a maximally-entangled state that is ``large enough" to provide them with $n$ images of the unknown state $S$. Then Alice can design her measurement such that \textit{for each outcome}: (1) a different part of $S$ is extracted from each image; and (2) the whole state is preserved across the $n$ images. Afterward, Bob can recover the state simply by rearranging the various parts, which he will be able to do once Alice informs him of the outcome of her measurement. Alice's measurement does not provide them with any information about the original state they are attempting to teleport, nor does Bob's rearrangement require that they know anything about it. In all cases, they remain ignorant of the state they are teleporting. The reader is encouraged to draw a diagram (perhaps for coins with $n=3$ sides each; see the following paragraph) and follow through the argument to be sure it is clear how this is done. The diagram will have $n\times n=n^2$ horizontal rows (representing Alice's two $n$-sided coins $A,a$, each row corresponding to one of the combinations of sides of these two coins: $HH,~HT,~HU,~HV,~\cdots$, where $H,T,U,V,\cdots$ label the various sides), and $n$ vertical columns (representing Bob's coin $b$). 

A complete measurement for the $n=3$ case will include $n^2=9$ outcomes, but an essentially complete understanding can be gained if the reader considers only the three outcomes corresponding to (1) $\langle H_AH_a| + \langle T_AT_a| + \langle U_AU_a|$; (2) $\langle H_AT_a| + \langle T_AU_a| + \langle U_AH_a|$; and (3) $\langle H_AU_a| + \langle T_AH_a| + \langle U_AT_a|$, where $U$ is the third side of these coins (the other six outcomes involve additional complications that I have not explained here, but these outcomes are not crucial for the general kind of understanding we are aiming for here). In this case the appropriate generalization of $B_0$ is the state $|H_aH_b\rangle + |T_aT_b\rangle + |U_aU_b\rangle$, and the three terms in this expression yield the three images needed for teleportation.

\section{Teleporting classical coins}
In this section, I consider teleportation of classical coins, which turns out to be possible using a method that bears a striking resemblance to the method used for quantum coins. Imagine that Chloe prepares a classical coin (labeled $A$) as either $H$ or $T$, and gives it to Alice, who is not allowed to look at the coin. Chloe also prepares classical coins $a$ and $b$ such that they are either $HH$ (both $H$) or $TT$ (both $T$). She then gives coin $a$ to Alice and coin $b$ to Bob, but again does not allow these parties to look at their coins. Alice now asks Chloe the following two questions: Are coins $A$ and $a$ the same? Or are they different? This pair of ``yes-no" questions represents a measurement, as defined earlier, on this pair of coins. If Chloe informs her they are the same, then Alice knows that coin $b$, which is guaranteed to be the same as $a$, is also the same as $A$; if, on the other hand, Chloe says coins $A$ and $a$ are different, then coin $b$ is also different from $A$. Alice now calls Bob on the telephone and tells him to ``flip" or ``don't flip". In the first case ($A$ same as $a$) she tells him not to flip, while when $A$ and $a$ are different, she tells him to flip. After he follows her instruction, Bob's coin $b$ will with certainty match coin $A$. The state of coin $A$ has been teleported onto coin $b$.

It is instructive to look at why the quantum case is astonishing while the classical one is rather mundane. There are three important differences between classical and quantum teleportation. The first difference has to do with the information that Alice would need to transmit to Bob in order to inform him of the state of coin $A$, if she happened to know that state. For a classical coin, there are only two possibilities, $H$ or $T$, so she would need to transmit only one bit to Bob. This is the same amount of information that is actually transmitted when she tells him ``flip" or ``don't flip" -- again, two possibilities. In contrast, as was discussed at the beginning of Section~\ref{vqip} for the case of quantum coins, it would require an \textit{infinite} amount of information for Alice to inform Bob of the state of coin $A$, whereas she only actually transmits two bits of information when informing him which of her four questions was the outcome of her measurement. We see that the two cases, classical vs. quantum, are dramatically different in terms of the amounts of information involved.

The second difference between these two cases is a bit more subtle. In the classical case, if Alice were to cheat and actually look at coin $A$, she would automatically know what state that coin is in and be able to tell Bob what to do with his coin -- turn it $H$ or turn it $T$; another one-bit message encompassing these two possibilities. This absolutely will not work for a quantum coin, which Alice cannot simply ``look" at to discover its state. The reason is the following: To begin with, in contrast to a classical coin, when Alice looks at her quantum coin, she invariably disturbs it in the process. That is, no matter what state the coin was in before she looked at it, the state after she looks at it is with certainty given by the outcome of her measurement. For example, even if the state is ``equal parts $H$ and $T$" before she asks if it is $H$ or $T$, if the answer is $H$ ($T$), then the coin is now $H$ ($T$). Or if it is $H$ to begin with and she does a measurement that answers yes to the question ``Is it equal parts $H$ and $T$?", then the state of the coin will now be equal parts $H$ and $T$. Hence, when she looks at it, she will get one of two answers (those being the two possible outcomes of her measurement) as to the state of the coin, but if she looks at it \textit{wrong}, that answer will not tell her \textit{what the state was beforehand}, but only what it is now. Furthermore, since she has now disturbed the state, there is no way to go back and try again, since the coin is now in a completely different state than the one she is trying to discover. The moral of this story is twofold: With quantum coins (1) don't bother trying to cheat; and (2) there's no point in asking for a ``do-over".

The third difference between the quantum and classical cases is even more subtle and is related to entanglement, for which there is no counterpart with classical coins. For classical teleportation, Chloe must tell Alice whether or not coins $a,A$ are the same or different. When these coins are classical, and since Chloe is the one that prepared them, she is certainly able to do so. However, in the quantum case coins $a,b$ are entangled, which means that neither one has a definite state of its own. Since coin $a$ does not have a definite state, the question whether coins $a,A$ are the same (have the same state) \textit{has no answer}! Even if we assume that Chloe prepared coins $a,b$ in their entangled state, there is nothing whatsoever that she (or anyone else) can say about the state of coin $a$, except that it is entangled with $b$ and has no definite state of its own. It is worth noting that in both the quantum and classical cases, coins $a,b$ are correlated with each other in ways that at first glance appear to be very similar -- when one is measured and found to be $H$ ($T$), the other one will also be $H$ ($T$). Nonetheless, the correlations present in the entangled state $B_0$ of quantum coins $a,b$ have \textit{no analog} in the case of classical coins. One reason is precisely what we have just discussed: that the quantum coins can be correlated in this way even though neither of the individual coins has a definite state of its own (a classical coin \textit{always} has a definite state of its own).

\section{Conclusion}
I have described a novel way to visualize the processing of quantum information, and used this picture to give a simple way to ``see" how teleportation is possible. The picture turns out to be useful beyond just providing an understanding of previously known phenomena (teleportation), however. Indeed, it has given us a deeper understanding of the process of deterministically implementing non-local unitaries by local operations and classical communication (when shared entanglement is available as a resource), allowing us to construct new protocols \cite{ourNLU} that go far beyond what was previously known to be possible \cite{ReznikStator}. We have also used this picture to study the question of what entanglement resources are required to locally implement other non-local operations, such as measurement protocols for the purpose of distinguishing sets of quantum states that are indistinguishable without the extra entangled resource \cite{inpressNote}.

\begin{acknowledgments}
This work has been supported in part by the National Science Foundation through Grant No. PHY-0456951. I am very grateful for numerous discussions with Bob Griffiths and others in his research group.
\end{acknowledgments}

\begin{thebibliography}{10}
\bibitem{minusNote}
The other two questions involve minus signs and are represented as (3) $\langle H_AH_a|-\langle T_AT_a|$ and (4) $\langle H_AT_a|-\langle T_AH_a|$, respectively. Multiplying one or the other of the states $H$ and $T$ by a minus sign is another valid quantum operation, which Bob can perform on his coin. By doing so, he effectively turns each of these cases into one of the two already shown explicitly in the paper. When question (3) is answered affirmatively, multiplication of $T_b$ by the minus sign is all that is needed to complete the teleportation; and for question (4), the minus sign operation need only be followed by Bob flipping his quantum coin.
\end{thebibliography}

\end{document}